\documentclass[12pt]{article}
\usepackage{amssymb,amsmath,epsfig,cite}

\normalbaselines

\begin{document}

\title{Wronskian formula for confluent second-order supersymmetric quantum 
mechanics}

\author{David J. Fern\'andez C.${}^\dagger$ and Encarnaci\'on 
Salinas-Hern\'andez${}^\ddagger$ \\ [5pt]
\small${}^\dagger$Departamento de F\'{\i}sica, CINVESTAV \\
\small A.P. 14-740, 07000 M\'exico D.F., Mexico \\ [5pt]
\small ${}^\ddagger$Escuela Superior de F\'{\i}sica y Matem\'aticas 
and Escuela Superior de C\'omputo \\ 
\small Instituto Polit\'ecnico Nacional, 
Ed. 9, U.P. Adolfo L\'opez Mateos \\ 
\small 07738 M\'exico D.F., Mexico}

\date{}

\maketitle

\begin{abstract}
The confluent second-order supersymmetric quantum mechanics, with 
factorization energies $\epsilon_1, \ \epsilon_2$ tending to a single 
$\epsilon$-value, is studied. We show that the Wronskian formula remains 
valid if generalized eigenfunctions are taken as seed solutions. The 
confluent algorithm is used to generate SUSY partners of the Coulomb 
potential.
\end{abstract}



\section{Introduction.}

The supersymmetric quantum mechanics (SUSY QM) represents a powerful tool 
for the spectral design in quantum theory. It has proved successful to 
construct solvable potentials for which the spectral information can be 
anallytically determined \cite{crf01,afhnns04,mr04,ac04,ff05}. The 
simplest version is the first-order SUSY QM which, however, does not allow 
to modify levels different from the ground state without creating 
singularities in the SUSY partner potential \cite{mr04}. As an alternative 
to surpass this problem, the second-order SUSY QM can be used 
\cite{ais93,bs97,fe97,bgbm99,mnr00,ast01,fmr03, 
fs03}. Indeed, through this technique one can embed two levels 
$\epsilon_1, \ \epsilon_2$ between two neighbouring energies of the 
initial Hamiltonian \cite{ff05,bs97}. It is possible as well to create 
single levels above $E_0$ \cite{fs03} and to generate complex potentials 
with either purely real spectra or having some complex `energies' 
\cite{fmr03}.

For $\epsilon_1 \neq \epsilon_2$ the modification to the potential induced 
by the second-order SUSY QM involves the Wronskian of the associated seed 
Schr\"odinger solutions $u_1, \ u_2$. In the confluent case 
$\epsilon_1=\epsilon_2\equiv\epsilon$, however, the key function $w(x)$ 
inducing the change depends just of one seed solution $u$ and it is not 
clear that the treatment based on the Wronskian remains valid. In this 
paper we will prove that result by identifying appropriate (generalized) 
eigenfunctions for which the Wronskian coincides with $w(x)$. This implies 
that a unified technique valid in the non-confluent and in the confluent 
case is available, we just have to identify the right seed solutions.

In the next section we will address the second-order SUSY QM and its 
classification scheme. We will study in some detail the confluent case, 
the restrictions onto the seed solution to ensure the regularity of the 
potentials difference and the discussion about the Wronskian. Then, we 
will generate confluent second-order SUSY partners of the Coulomb 
potential (some of which are new). We will end the paper with our 
conclusions.

\section{Second-order supersymmetric quantum mechanics.}

The second-order SUSY QM consists in the following realization of the 
standard SUSY algebra with two generators:
\begin{eqnarray}
&&\{ Q_j,Q_k\} = \delta_{jk} H_{ss}, \quad [H_{ss},Q_j] =  0,
\quad j,k  =  1,2 \\
&& Q_1 =\frac{Q^\dagger + Q}{\sqrt{2}},
\quad Q_2 = \frac{Q^\dagger - Q}{i\sqrt{2}}, \quad
Q = \left(\begin{matrix}
0 & A \cr 0 & 0
\end{matrix}\right), \quad 
Q^\dagger = \left(\begin{matrix}
0 & 0 \cr A^\dagger & 0
\end{matrix}\right)\\
&& H_{ss}  = 
\left(\begin{matrix}
A A^\dagger & 0 \cr 0 & A^\dagger A
\end{matrix}
\right)  =  \prod_{j=1}^2 \left(
\begin{matrix}
\widetilde H - \epsilon_j & 0 \cr 0 & H - \epsilon_j
\end{matrix}
\right) \label{factorized} \\
&& H = - \frac{d^2}{dx^2} + V(x), \quad 
\widetilde H = -\frac{d^2}{dx^2} + \widetilde V(x)
\label{defHHtil} \\
&& A = \frac{d^2}{dx^2} + \eta(x)\frac{d}{dx} + \gamma(x), \quad
\widetilde HA = AH  \label{intertwining}
\end{eqnarray}
The intertwining relationship (\ref{intertwining}) leads to the following 
set of equations:
\begin{eqnarray}
&& \frac{\eta \eta^{\prime\prime}}{2}-\frac{{\eta^{\prime}}^2}4
+\eta^2\bigg(\frac{\eta^2}4 -\eta^{\prime} - V + d\bigg) + c=0
\label{eqeta}  \\
&& \gamma = d - V +\frac{\eta^2}2 - \frac{\eta^\prime}2,
\qquad \widetilde V = V + 2 \eta'  \label{newV} 
\end{eqnarray}
Given $V(x)$, $c$ and $d$, the new potential $\widetilde V(x)$ is 
determined by the solutions $\eta$ to the non-linear equation 
(\ref{eqeta}). To find them, let us use the Ans\"atz:
\begin{eqnarray}
&& \eta'(x) = \eta^2(x) + 2 \beta(x) \eta(x) - 2\xi(x) \label{ansatz}
\end{eqnarray}
which leads to $\xi^2 =c$ and the following Riccati equation:
\begin{eqnarray}
& \beta^\prime+\beta^2=V-\epsilon, \quad \epsilon = d + \xi 
\label{riccati}
\end{eqnarray}
We can work instead with the equivalent Schr\"odinger equation ($\beta = 
u'/u$)
\begin{eqnarray}
&& - u'' + V u = \epsilon u \label{schro}
\end{eqnarray}
Thus, the solutions $\eta$ can be classified according to the sign of $c$.

\section{Classification of second-order SUSY transformations.}

\subsection{The non-confluent cases with $c\neq 0$.} 

Here $\epsilon_1 \equiv d + \sqrt{c}, \ \epsilon_2 \equiv d - \sqrt{c}$, 
$\epsilon_1 \neq \epsilon_2$; this includes the real case with $c>0, \ 
\epsilon_1,\epsilon_2 \in {\mathbb R}$ and the complex one with $c<0, \ 
\epsilon_1 \in {\mathbb C},\epsilon_2=\epsilon_1^*$. From (\ref{ansatz}) 
two equations for $\eta$ are obtained:
\begin{eqnarray*}
& \eta' = \eta^2 + 2 \beta_1 \eta - (\epsilon_1 - \epsilon_2), \quad
& \eta' = \eta^2 + 2 \beta_2 \eta + (\epsilon_1 - \epsilon_2)
\end{eqnarray*}
which leads to:
\begin{eqnarray}
&& \eta =  \frac{\epsilon_1 - \epsilon_2}{\beta_1 -
\beta_2} = - \frac{d}{dx}\ln[W(u_1,u_2)]  \label{etanonconf}
\end{eqnarray}
where $W(f,g) = fg' - gf'$ is the Wronskian of $f$ and $g$. To avoid 
singularities in $\eta(x)$ the Wronskian has to be nodeless. The spectrum 
of $\widetilde H$, ${\rm Sp}(\widetilde H)$, will depend as well on the 
normalizability of the two eigenfunctions $\widetilde\psi_{\epsilon_j}$ of 
$\widetilde H$ with eigenvalue $\epsilon_j$ in the Kernel of $A^\dagger$, 
with explicit expressions given by $\widetilde\psi_{\epsilon_1}\propto 
u_2/W(u_1,u_2), \ \widetilde\psi_{\epsilon_2}\propto u_1/W(u_1,u_2)$.

Some spectral design possibilities are worth to be mentioned (for a 
detailed treatment see \cite{ff05}). {\it i)} Two levels $\epsilon_1, \ 
\epsilon_2$ can be created {\it below} the ground state energy of $H$, 
namely, $\epsilon_1 <\epsilon_2 < E_0$, ${\rm Sp}(\widetilde H) = 
\{\epsilon_1,\epsilon_2,E_n, n=0,1,2,\cdots\}$ \cite{fe97}. {\it ii)} A 
pair of levels can be placed {\it between} two neighbouring energies of 
$H$, i.e., $E_i<\epsilon_1<\epsilon_2<E_{i+1}$, ${\rm Sp}(\widetilde H) = 
\{E_0, \cdots,E_{i},\epsilon_1,\epsilon_2, E_{i+1},\cdots\}$ \cite{bs97}. 
{\it iii)} Two neighbouring energies of $H$ can be {\it deleted}, namely, 
$\epsilon_1 = E_i$, $\epsilon_2 = E_{i+1}$, ${\rm Sp}(\widetilde H) = 
\{E_0,\cdots,$ $E_{i-1}, E_{i+2},\cdots\}$ \cite{ff05}. {\it iv)} Some 
{\it complex} energies can be manufactured (the new Hamiltonians are 
non-hermitian) \cite{fmr03}.

Cases {\it ii), iii), iv)} run against the dominant idea that in SUSY QM 
the new levels are always real and {\it below} the ground state energy of 
the initial Hamiltonian, improving thus our spectral design possibilities 
\cite{ac04,ff05}.

\bigskip

\subsection{The confluent case with $c = 0$.} 

We have now that $\epsilon \equiv\epsilon_1 = \epsilon_2$, and the 
Ans\"atz (\ref{ansatz}) just provides \cite{mnr00,fs03}:
\begin{eqnarray*}
&& \eta' = \eta^2 + 2 \beta \eta  
\end{eqnarray*}
This Bernoulli equation has a general solution given by
\begin{eqnarray}
&& \eta(x) = \frac{e^{2\int\beta(x)dx}}{w_0 -
\int e^{2\int\beta(x)dx}dx} 
= - \frac{d}{dx}\ln[w(x)] \label{etaconf}
\end{eqnarray}
where, up to an unimportant constant factor, 
\begin{eqnarray}
&& w(x) = w_0 - \int_0^x u^2(y)dy  \label{wconf}
\end{eqnarray}
Since equations (\ref{etanonconf},\ref{etaconf}) look similar, perhaps 
$w(x)$ has to do with a Wronskian. To see that, suppose there is a 
function $v$ related with the given $u$ through:
\begin{eqnarray}
(H-\epsilon)v = u, \quad (H-\epsilon)u=0
\label{jsdiff}
\end{eqnarray}
i.e., $u$ and $v$ are rank $1$ and rank $2$ generalized eigenfunctions of 
$H$. This fact was noticed in \cite{ac04}, but a further study exploring 
the link with the Wronskian formula is needed. Let us find now $v$ from 
the differential equation (\ref{jsdiff}):
\begin{eqnarray}
&& v = u\left(k + \int \frac{w(x)}{u^2(x)} dx\right)
\end{eqnarray}  
Up to a constant factor, it turns out that $w(x) = W(u,v)$.

Similarly as for $\epsilon_1 \neq \epsilon_2$, the regular confluent 
transformation will be produced by a nodeless $w(x)$. If the domain of $x$ 
is ${\mathbb R}$, it is sufficient to use solutions $u(x)$ vanishing 
either when $x\rightarrow -\infty$ or when $x\rightarrow\infty$ 
\cite{fs03}. However, if the $x$-domain is the positive semi-axis, as in 
the Coulomb problem which we will address later, the simplest choice is to 
take:
\begin{eqnarray}
&& \lim_{x\rightarrow 0} u(x) = 0 \quad {\rm or} \quad
\lim_{x\rightarrow \infty} u(x) = 0   \label{uvp}
\end{eqnarray}
For $\epsilon\not\in {\rm Sp}(H)$ one of the requirements (\ref{uvp}) can 
be achieved, but not both simultaneously. For $\epsilon\in {\rm Sp}(H)$ 
the two conditons (\ref{uvp}) are automatically satisfied. In both cases a 
normalized eigenfunction of $\widetilde H$ with eigenvalue $\epsilon$ in 
the Kernel of $A^\dagger$ can be found, $\widetilde\psi_{\epsilon} \propto 
u/w$. In particular, for $\epsilon\geq E_0$ equation (\ref{uvp}) can be 
fulfilled, i.e., one level can be created above $E_0$. Let us remind that 
the excited state solutions obey (\ref{uvp}) and thus they are appropriate 
to generate new potentials through the confluent SUSY algorithm.

\section{Confluent SUSY partners of the Coulomb potential.}

Let us apply the confluent algorithm to the Coulomb problem which, after 
separating the angular variables and taking $\hbar=e=m=1$, leads to a 
Schr\"odinger Hamiltonian as given in (\ref{defHHtil}), with an effective 
potential
\begin{eqnarray}
&& V(r) = -\frac{2}{r} + \frac{\ell(\ell+1)}{r^2} \label{coulomb}
\end{eqnarray}
where $\ell=0,1,\dots$. The spectrum of the corresponding Hamiltonian 
reads ${\rm Sp}(H) = \{E_n = -1/n^2, \ n=\ell + 1,\ell + 2,\dots\}$. First 
we find the general solution to (\ref{schro}) with the $V(r)$ of 
(\ref{coulomb}) and an arbitrary factorization energy $\epsilon<0$; 
second, the solution with the right behaviour is chosen (see 
Eq.~(\ref{uvp})). After implementing the procedure, the solution vanishing 
at the origin reads:
\begin{eqnarray}
& u(r) = 
\sqrt{\frac{(-\epsilon)\Gamma\left(\ell+1+\frac{1}{\sqrt{-\epsilon}} 
\right)}{\Gamma\left( 
\frac{1}{\sqrt{-\epsilon}} - \ell\right)[\Gamma(2\ell + 2)]^2} }
\left(2r\sqrt{-\epsilon}\right)^{\ell + 1} e^{-r\sqrt{-\epsilon}} \ 
{}_1F_1\left(\ell + 1 
-\frac{1}{\sqrt{-\epsilon}},2\ell + 2;2r\sqrt{-\epsilon}\right) 
\label{ucoulomb}
\end{eqnarray}
$\Gamma(z)$ and ${}_1 F_1(a,b;z)$ being the Gamma and the confluent 
hypergeometric functions respectively. Notice that $u(r)$ becomes the 
normalized eigenfunction of $H$ for $\epsilon=E_n$. A straightforward 
calculation leads now to our key $w$-function:
\begin{eqnarray}
w(r)  & = &
w_0 \nonumber  - 
\sum_{m=0}^\infty \frac{\sqrt{-\epsilon}\, 
B\left(\ell + 1 + \frac{1}{\sqrt{-\epsilon}},\ell + 1 + m 
- \frac{1}{\sqrt{-\epsilon}}\right) 
(2r\sqrt{-\epsilon})^{2\ell + m + 3}}{2(2\ell + m + 3)(2\ell + 1)!\,m!\, 
B\left(\frac{1}{\sqrt{-\epsilon}} - \ell,\ell 
+ 1 - \frac{1}{\sqrt{-\epsilon}} 
\right)} \nonumber \\ &&
\hskip1cm 
\times \ 
{}_2F_2\left(2\ell + m + 3,\ell + 1 
+ \frac{1}{\sqrt{-\epsilon}};2\ell + m + 
4,2\ell + 2;-2r\sqrt{-\epsilon}\right) 
\label{wcoulombinf}
\end{eqnarray}
where $B(x,y)$, ${}_2F_2(a_1,a_2;b_1,b_2;z)$ are the Beta and a 
generalized hypergeometric function. For $\epsilon=E_n=-1/n^2$ this 
infinite series truncates:
\begin{eqnarray}
w(r)  & = & w_0 -
\sum_{m=0}^{n - \ell - 1}\frac{(-1)^m\,
(\frac{2r}{n})^{2\ell + m + 
3}}{(2\ell + m + 3)(2\ell + m + 1)} \nonumber \\ && \hskip1cm
\times \ \frac{{}_2F_2(2\ell + m + 3,n + \ell + 1;2\ell + m + 
4,2\ell + 2;-\frac{2r}{n})}{2n(2\ell + 1)!m!B(n - \ell - 
m,2\ell + m + 1)}
\label{wcoulombfin}
\end{eqnarray}
The $w_0$-domain where $w(r)$ is nodeless reads $(-\infty,0]$ for 
$\epsilon\neq E_n$ and $(-\infty,0]\cup[1,\infty)$ for $\epsilon = E_n$.

The SUSY partner potentials take the form:
\begin{eqnarray}
\widetilde V(r) = -\frac{2}{r} + 
\frac{\ell(\ell + 1)}{r^2} + \frac{4u(r)u'(r)}{w(r)} +  
\frac{2u^4(r)}{w^2(r)} \label{nvuw}
\end{eqnarray}
In (\ref{nvuw}) $u(r)$, $w(r)$ are given by (\ref{ucoulomb}) and 
(\ref{wcoulombinf},\ref{wcoulombfin}) while $u'(r)$ arises from 
(\ref{ucoulomb}). Some examples, induced by the $w(r)$ of 
(\ref{wcoulombfin}), deserve an explicit discussion.

\subsection{The case with $\ell=0$ and $n=1$.}

Here the SUSY partner potentials of $V(r) = -2/r$ are given by:
\begin{eqnarray}
&& \widetilde V(r) = - \frac{2}{r} - 
\frac{16\,r\,\left[ - 1 - r + (w_0 - 1)(r - 1)e^{2r} \right]}
{{\left[ 1 + 2r + 2r^2 + (w_0 - 1)e^{2r}\right]}^2} \label{nvl0n1}
\end{eqnarray}
The potentials $V(r), \ \widetilde V(r)$ are isospectral for $w_0\in 
(-\infty,0)\cup (1,\infty)$ and they differ in the ground state for $w_0 = 
0,1$ because $E_1=-1$ is missing of $\widetilde H$ in the last case. The 
potentials (\ref{nvl0n1}) coincide with a family derived long ago through 
the factorization method, we just take $w_0 = 4 \gamma_1$ in 
(\ref{nvl0n1}) to get equation (3.1) of \cite{fe84} (see also 
\cite{fmr03,ad88,ro98}).

\subsection{The case with $\ell=0$ and $n=2$.}

The SUSY partner potentials of $V(r) = -2/r$ become now
\begin{equation}
\widetilde V(r) = - \frac{2}{r} +
\frac{8r(r - 2)\left[- 8 + 4r + 6r^2 + 2r^3 + r^4 - 
2(w_0 - 1)\left(4 - 6r + r^2 \right)e^r\right]}
{{\left[ 8 + 8r + 4r^2 + r^4 + 8(w_0 - 1)e^r\right] }^2}
\label{nvl0n2}
\end{equation}
Once again, $V(r)$ and $\widetilde V(r)$ are isospectral for $w_0\in 
(-\infty,0)\cup (1,\infty)$, and for $w_0 = 0,1$ their spectra differ 
because the level $E_2=-1/4$ is missing of $\widetilde H$. As far as we 
know, the potentials (\ref{nvl0n2}) have not been derived previously.

Other families of exactly solvable potentials can be written explicitly 
for different values of $n$ and $\ell$. We illustrate one of them in 
figure 1 for $n=4$, $\ell=1$ and $w_0=-0.1$. For comparison, the initial 
potential is also shown.

\begin{figure}[t]
\begin{center}
\epsfig{file=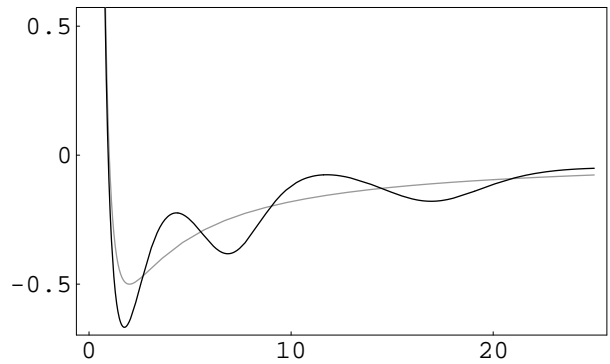, width=8cm}
\caption{Isospectral SUSY partner potentials $\widetilde V(x)$ (black 
curve) and $V(r) = -2/r$ (gray curve) generated through (\ref{ucoulomb}), 
(\ref{wcoulombfin},\ref{nvuw}) with $n=4$, $\ell=1$, $w_0=-0.1$.
\label{fig1}}
\end{center}
\end{figure}

\section{Conclusions}

Contrasting with the first-order SUSY QM, in which we modify just the 
ground state energy, the confluent algorithm allows the embedding of 
levels at any positions on the energy axis. We have shown that the 
Wronskian formula is still valid in the confluent case. We conclude that 
the second-order SUSY QM represents a poweful tool in which the right 
choice of the seed solutions allows us to generate the new potential in a 
simple way.

\section*{Acknowledgements}

The authors acknowledge the support of Conacyt (M\'exico), project No. 
40888-F. ESH also acknowledges the support of the {\it Instituto 
Polit\'ecnico Nacional}.

\end{document}